\documentstyle{article}
\newcommand{\be}{\begin{equation}}
\newcommand{\ee}{\end{equation}}
\newcommand{\bea}{\begin{eqnarray}}
\newcommand{\eea}{\end{eqnarray}}
\newcommand{\bc}{\begin{center}}
\newcommand{\ec}{\end{center}}

\newcommand{\talkauthors}[1]{
{\small {\bf $\copyright$ \enskip 1993 \enskip \enskip #1}}}
\newcommand{\talktitle}[1]{{\small {\bf #1}}}
\newcommand{\address}[1]{{\small {\it #1}}}

\begin{document}
\begin{center}
\talkauthors{KORENBLIT S.E. } \\ [2.0mm]
\talktitle{ THE FROISSART-GRIBOV REPRESENTATION  \\
    FOR JOST FUNCTIONS OF DIRAC OPERATOR \\
             IN ARBITRARY DIMENSION SPACE} \\ [2.0mm]
\address{INSTITUTE OF APPLIED PHYSICS, IRKUTSK STATE UNIVERSITY,\\
          RUSSIA} \\
\end{center}
\vskip 3mm
\begin{center} Abstract \end{center}

{\small
   A dynamic scheme basing on equation for T-matrix momentum
transfer spectral density and integral representation for Jost
function is proposed for local Dirac Hamiltonians in arbitrary N-
dimension spaces and for Schrodinger one with singular or nonlocal
generalized Yukawa-type potentials.}

 A generalization of the off-shell-Jost function method for
that Ha\-mil\-to\-ni\-ans and universal renormalization procedure of
Jost function calculation for singular and nonlocal potentials
is proposed.

\section{\rm Introduction}

    It is well known, that determinant
\be
{\bf d}(W)=det\left[{\bf G}_{0}(W) {\bf G}_{V}^{-1}(W) \right]=
det\left[{\bf I}-{\bf G}_{0}(W) V\right]
=det\left[{\bf I}+{\bf G}_{V}(W) V \right]^{-1},
\label{1}
\ee
with Green function (resolvent) ${\bf G}_{V}(W)=\left[W - H_{V}\right]^{-1}$,
accumulate all observable information about spectra of the stationary
Hamiltonian $H_{V} = H_{0} + V$ in most economical form \cite{BW}:
\begin{equation}
{\bf d}(W)=
\prod_{n=1}^{n^{max}}\biggl(1-\frac{W_{n}}{W}\biggl)
exp
\left\{-
\frac{1}{\pi}\int_{0}^{\infty}
d \varepsilon
\frac{\delta (\varepsilon)}{(\varepsilon-W)}
\right\}
\label{2}
\end{equation}
which makes it very convenient for solving both direct and inverse scattering
problems \cite{AR,N,CS}, and for finding a different sums on
spectra $H_{V}$ \cite{Krz}. It arise also in one-loop calculations for
different quantum effects in external fields $V(\vec{\rm x})$ or in the
semiclassical quantization of field theory near nontrivial classical
solutions \cite{IZ,Br,CC}. However, derivation of this determinant
usually imply solution of two different eigenvalue problems for finding
characteristics of discrete $W_{n}$ and continuous $\delta(\varepsilon)$
spectra separately. That makes their calculation and utilization much more
complicated.
This circumstance stimulates search of another ways for construction
determinant does not requiring any information about $H_{V}$ eigenvalues.

On the other hand, in the case of spherically symmetrical Hamiltonian
$V(\vec{\rm x})=V(r) $, using Green
function's partial expansion onto irreducible representations of rotation
group SO(N), for instance, for Dirac operator:
\be
\begin{array}{l}
\!\!\!\!\displaystyle
<\vec{\rm x}|{\bf G}^{(N)}_{V}(W^{\overline{\zeta}}(ib))|\vec{\rm y}>=
\frac{1}{(ry)^{(N-1)/2}}\cdot \\[4mm]
\displaystyle \cdot
\sum^{\infty}_{J_{\rm N}=\lambda_{N}}
\sum_{\xi=\pm 1}
\left(
\begin{array}{cc}
\Pi_{\kappa_{\xi}}(\vec{\rm n},\vec{\omega})\
G^{\overline{\zeta}\, 11}_{\kappa_{\xi} V}
&
i \Pi_{\kappa_{\xi}}(\vec{\rm n},\vec{\omega})\,
(\vec{\sigma}\vec{\omega})\
G^{\overline{\zeta}\, 12}_{\kappa_{\xi} V}
\nonumber \\
-i (\vec{\sigma} \vec{\rm n})^{\dagger}\,
\Pi_{\kappa_{\xi}}(\vec{\rm n},\vec{\omega})\
G^{\overline{\zeta}\, 21}_{\kappa_{\xi} V}
&
(\vec{\sigma}\vec{\rm n})^{\dagger}\,
\Pi_{\kappa_{\xi}}(\vec{\rm n},\vec{\omega})\,
(\vec{\sigma} \vec{\omega})\
G^{\overline{\zeta}\, 22}_{\kappa_{\xi} V}
\\
\end{array}
\right)
\end{array}
\label{3}
\ee
where: $ \vec{\rm x}=r \vec{\rm n};\ \vec{\rm y}=y\vec{\omega}$, and
$\Pi_{\kappa_{\xi}}(\vec{\rm n},\vec{\omega}) $ is projector onto subspace
with fixed orbital $l^{(N)}_{\xi}$ and total $J_{\rm N}$ angular momentum
(see Appendix);
one can formally factorize ${\bf d}(W)$ into infinite product:
\begin{equation}
{\bf d}^{(N)}_{Dir}(W^{\overline{\zeta}}(ib))=
\prod^{\infty}_{J_{\rm N}=\lambda_{N}}
\prod_{\xi=\pm 1}
\left[
F^{\overline{\zeta}}_{\kappa_{\xi}}(b)\right]^{\Delta(N,{J_{\rm N})}}
\label{4}
\end{equation}
where dimension of the SO(N)-representation for this case is
\footnote {The minimal gamma-matrix representation is assumed.}:
\begin{equation}
Tr\{\Pi_{\kappa_{\xi}}\}=\Delta(N,J_{\rm N})=
2^{[(N-1)/2]}\frac{(J_{\rm N}+\lambda_{N})!}
{(J_{\rm N}-\lambda_{N})!(N-2)!};
\label{5}
\end{equation}
and the following notations are accepted hereafter:
\begin{eqnarray}
&& a_{N}=\frac{1}{2}(3-N);\quad
\lambda_{N}=\frac{1}{2}-a_{N}=\frac{N}{2}-1;\quad
\kappa \equiv \kappa_{\xi}=\xi \left(J_{\rm N}+\frac{1}{2}\right);
\nonumber \\
&& L_{\xi}=J_{\rm N}+\frac{\xi}{2}=l^{(N)}_{\xi}-a_{N}; \qquad
\xi=\pm 1.
\label{6}
\end{eqnarray}
with the numbers $ l^{(N)}_{\xi} = 0,1,2,.., $ and $ J_{\rm N}=
\lambda_{N},\lambda_{N}+1,\lambda_{N}+2,... $
defining eigenvalues of squared orbital and squared total angular
momentum respectively \cite{Jh} (see Appendix)
\[
\frac{1}{2}({\bf L \cdot L}) \Rightarrow
l_{\xi}^{(N)}\left(l_{\xi}^{(N)}+2\lambda_{N}\right);\
\frac{1}{2}({\bf J \cdot J}) \Rightarrow \left(J_{\rm N}+
\frac{1}{2}\right)^{2}-\frac{1}{8}(N-1)(N-2).
\]
The partial determinants or Jost functions
$ F^{\overline{\zeta}}_{\kappa_{\xi}}(b) $
are defined by the same formulae (\ref{1}),(\ref{2})
with partial Green function which matrix elements are
$ G^{\overline{\zeta}\, ij}_{\kappa_{\xi} V}(b;r,y) $,
and with scattering phase $ \delta_{\kappa_{\xi}}(\varepsilon) $
\cite{Kl} respectively. In contrast
with ${\bf d}(W)$ \cite{BW}, they are well defined for arbitrary
local potential which is less singular at $r=0$, than appropriate free
Hamiltonian $H_{0}$, and disappear sufficiently fast at
$r\rightarrow \infty$ \cite{AR,N}.

There is still one problem on this way, of finding an integral
representation determined a general form of Jost function's
$l_{\xi}^{(N)}, J_{\rm N}$ - dependence. Such representation may be useful
both for field theoretical calculations mentioned above and for Regge
phenomenology  \cite{Koll}. There was an attempt made in \cite{ARR} for
Schrodinger case with N=3. It led to representation with two variable's
weight function which satisfies to complicate nonlinear integral equation
and does not have any known physical meaning \cite{AR,ARR}.

A quite different integral representation for Jost function (matrix) was
established recently in \cite{Kpr,Ktmf,Krc} for the Dirac operator
with N=3, and for the Schrodinger one in arbitrary N-dimension space and
in model with N strongly coupled channels.
It play a role, analogous to Froissart-Gribov representation for
partial amplitudes, but define the Jost functions in all analytical
region over complex variables $J_{\rm N}$ and $b$ in terms of quadratures
from half-off-shell T-matrix spectral density over momentum transfer, with
energetic variables, analytically continued from the continuum to the
bound state region, and provides a group-theoretical interpretation
directly for the Jost function. Together with linear Volterra-type integral
equation for the spectral density this representation forms a dynamic scheme,
from which all Jost functions (matrices) are found via solution of one
regular problem, which has nothing to do with eigenvalue one.

Present work gives a generalization of this scheme for a wide class of
operators, including N-dimension regular Dirac operator, singular or
nonlocal Schrodinger operators, and the last with relativistic corrections
to potential.

\section{\rm Equation for spectral densities}

The aim of this section is to derive equations for T-matrix
spectral densities over momentum transfer, constituting the
foundation for the dynamic scheme in question, and to elucidate
their analytical properties.

We define a family of Dirac operators in $R_{N}$: $ \mu,\nu=0,1,2,...N,$
\begin{equation}
H_{0}=(\vec{\Gamma} \cdot \vec{\bf P})+\Gamma_{0}m;\quad
\left\{
\Gamma_{\mu}\,,\Gamma_{\nu}
\right\}
=2\delta_{\mu\nu};\quad
\vec{\bf P}=-i\vec{\nabla}_{N};
\label{7}
\end{equation}
\begin{equation}
a) H_{V}=H_{0}+{\bf I}V; \qquad b) H_{V}=H_{0}+\Gamma_{0}V;
\label{8}
\end{equation}
with local Yukawa-type potentials
\begin{equation}
V(r)=\frac{4\pi}{\Omega_{N}\pi^{a}r}
\int_{\mu_{0}}^{\infty}d\nu\ \Sigma^{(N)}(\nu)\,
\left(
\frac{r}{2\nu}
\right)^{a}\, \chi_{-a}(\nu r),
\label{9}
\end{equation}
or in momentum representation \cite{Gf-Sh}
\footnote{Here subtractions lead to ultralocal terms
$(\Delta)^{n}\delta_{N}(\vec{\rm x})$ in (\ref{9}) corresponding to
regularization of the potential in the sense of distributions. For Dirac
Hamiltonian such singular potential is unstable with respect to particle
creation, and we assume the absence of subtractions for that case.}:
\begin{equation}
<\vec{\rm q}|V|\vec{\rm p}>=\frac{2}{\pi\Omega_{N}}
\int_{\mu_{0}}^{\infty}d\nu\frac{\Sigma^{(N)}(\nu)}{[\nu^{2}+
(\vec{\rm q}-\vec{\rm p})^{2}]}+
({\it subtractions}).
\label{10}
\end{equation}
The normalization conditions are:
\[
<\vec{\rm q}|\vec{\rm p}>=\delta_{N}(\vec{\rm q}-\vec{\rm p});\
<\vec{\rm x}|\vec{\rm p}>=exp(-\frac{1}{2} i \pi a_{N}) e^{i(\vec{\rm p}
\cdot \vec{\rm x})}(2\pi)^{-N/2};
\]
and the following notations are used hereafter:
$
\Omega_{N}=2\pi^{N/2}/\Gamma (N/2); \\ N \geq 2;\
\vec{\rm q}=q\vec{\tau},\ \vec{\rm p}=p \vec{\rm v};
$
\begin{equation}
\chi_{l}(\beta r)=(\frac{2}{\pi}\beta r)^{1/2} K_{l+\frac{1}{2}}(\beta r);
\qquad
\chi_{0}(\beta r)=e^{-\beta r};
\label{12}
\end{equation}
where $K_{m}(z)$ are McDonald function \cite{Bt-Er}.
Choosing for $\Gamma$-matrices the following representation \cite{NeSf}:
\be
(\vec{\Gamma})_{k}\equiv\Gamma_{k}=
\left(
\begin{array}{cc}
O & \sigma_{k}\\
\sigma^{\dagger}_{k} & O\\
\end{array}
\right)
,\qquad
\Gamma_{0}=
\left(
\begin{array}{cc}
{\tt I} & O\\
O & -{\tt I}\\
\end{array}
\right);
\label{13}
\ee
where matrices $\sigma_{k}$ for $k,j=1,2,...N $ satisfy to conditions:
\begin{equation}
\sigma_{j}\sigma^{\dagger}_{k}+\sigma_{k}\sigma^{\dagger}_{j}=
\sigma^{\dagger}_{j}\sigma_{k}+\sigma^{\dagger}_{k}\sigma_{j}=
2\delta_{jk},
\label{14}
\end{equation}
we have a complete set of eigenfunctions for operator (\ref{7}):
\[
H_{0}(\vec{\rm p})\ {\bf u}_{\zeta}(\vec{\rm p},[\lambda])=
w^{\zeta}(p)\ {\bf u}_{\zeta}(\vec{\rm p},[\lambda]);
\]
\[
{\bf u}_{\zeta}(\vec{\rm p},[\lambda])=
\left(
\begin{array}{c}
\zeta\sqrt{\varepsilon(p)+m\zeta}
\\
\sqrt{\varepsilon(p)-m\zeta} \quad (\vec{\sigma}\vec{\rm v})^{\dagger}
\end{array}
\right)
{\bf w}_{[\lambda]}(\vec{\rm v});
\]
\[
\sum_{[\lambda]}
{\bf u}_{\zeta}(\vec{\rm p},[\lambda])
\otimes
{\bf u}^{\dagger}_{\zeta}(\vec{\rm p},[\lambda])=
\varepsilon(p)+\zeta H_{0}(\vec{\rm p});
\]
\begin{equation}
\left(
{\bf u}^{\dagger}_{\zeta^{"}}(\vec{\rm p},[\mu])\cdot
{\bf u}_{\zeta}(\vec{\rm p},[\lambda])
\right)
=2\varepsilon(p)\ \delta_{\zeta^{"}\zeta}\ \delta_{[\mu][\lambda]},
\label{15}
\end{equation}
where following definitions are used:
\begin{equation}
\varepsilon(p)=+\sqrt{p^{2}+m^{2}};\ w^{\zeta}(p)=\zeta\varepsilon(p);\
W^{\overline{\zeta}}(ib)=\overline{\zeta}\sqrt{m^{2}-b^{2}};
\
\overline{\zeta},\zeta=\pm 1.
\label{16}
\end{equation}
Spinors ${\bf w}_{[\lambda]}(\vec{\rm n})$ with quantum numbers
$[\lambda]$ on
group SO(N) realize its spinor representation of half dimension than
${\bf u}_{\zeta}(\vec{\rm p},[\lambda])$, and satisfy to the following
conditions:
\begin{equation}
\left(
{\bf w}^{\dagger}_{[\mu]}(\vec{\rm n})\cdot
{\bf w}_{[\lambda]}(\vec{\rm n})
\right)
=\delta_{[\mu],[\lambda]};\qquad
\sum_{[\lambda]}
{\bf w}_{[\lambda]}(\vec{\rm n})
\otimes
{\bf w}^{\dagger}_{[\lambda]}(\vec{\rm n})={\tt I}
\label{17}
\end{equation}

We consider also Schrodinger operators for N=3 with relativistic
correction to potential $V(r)$ (here $\sigma_{1,2,3}$ are Pauly matrices):
\begin{equation}
H_{V}=\vec{\bf P}^{2}(2m)^{-1}+ {\bf U}(\vec{\rm x}),
\end{equation}
\begin{equation}
{\bf U}(\vec{\rm x})=V(r)-\frac{1}{2}(2m)^{-1}
\left[
(\vec{\sigma} \cdot \vec{\bf P}),
\left[
(\vec{\sigma} \cdot \vec{\bf P}),V(r)
\right]
\right]
\label{18}
\end{equation}
and with nonlocal interaction:
\begin{equation}
{\bf U}(\vec{\rm x})=V_{1}(r)+
(2m)^{-2}(\vec{\bf P}^{2}V_{2}(r)+V_{2}(r)\vec{\bf P}^{2}).
\quad
\label{19}
\end{equation}

Using definitions (\ref{15}) and the Lippman-Schwinger (LS) equation
\be
{\bf T}(W) = {\bf V} + {\bf V}{\bf G}^{c}_{0}(W) {\bf T}(W),
\label{20}
\ee
with the help of free Green function's decomposition:
\begin{eqnarray}
{\bf G}^{c}_{0}(W^{\overline{\zeta}};k)
& = &
(W^{\overline{\zeta}}+H_{0}(\vec{\rm k}))
\left[
(W^{\overline{\zeta}})^{2}-\vec{\rm k^{2}}-m^{2}-i0
\right]^{-1}=
\nonumber \\
& = &
\frac{1}{2\varepsilon(k)}
\sum_{\zeta^{'}=\pm 1}
\sum_{[\lambda]}
\frac{
{\bf u}_{\zeta^{'}}(\vec{\rm k},[\lambda])
\otimes
{\bf u}^{\dagger}_{\zeta^{'}}(\vec{\rm k},[\lambda])
     }
{
\left[
W^{\overline{\zeta}}-\zeta^{'}(\varepsilon(k)-i0)
\right]
},
\nonumber
\end{eqnarray}
we define for Hamiltonian (\ref{7}),(\ref{8}) T-operator acting on
spinors (\ref{17}):
\be
{\bf w}^{\dagger}_{[\mu]}(\vec{\tau})\
{\bf(}\vec{\rm q},\zeta^{"}|
{\bf T}(W^{\overline{\zeta}})|
\vec{\rm p},\zeta{\bf )}\
{\bf w}_{[\lambda]}(\vec{v})
={\bf u}^{\dagger}_{\zeta^{"}}(\vec{\rm q},[\mu])
<\vec{\rm q}|
{\bf T}(W^{\overline{\zeta}})|
\vec{\rm p}>
{\bf u}_{\zeta}(\vec{\rm p},[\lambda]),
\label{21*}
\ee
with symmetry properties:
\be
{\bf (}\vec{\rm q},\zeta^{"}|
{\bf T}(W^{\overline{\zeta}})|
\vec{\rm p},\zeta{\bf )}=
{\bf (}-\vec{\rm q},\zeta^{"}|
{\bf T}(W^{\overline{\zeta}})|
-\vec{\rm p},\zeta{\bf )}
=\left(
{\bf (}\vec{\rm p},\zeta|
{\bf T}(W^{\overline{\zeta}})|
\vec{\rm q},\zeta^{"}{\bf )}
\right)^{\dagger}.
\label{21}
\ee
It is not difficult to see from (\ref{10}),(\ref{20}), that it possess
spectral rep\-re\-sen\-ta\-tion:
\begin{eqnarray}
& &
{\bf (}\vec{\rm q},\zeta^{"}|\stackrel{(N)}{\bf T}
(W^{\overline{\zeta}}(ib))|
\vec{\rm p},\zeta{\bf )}=
\frac{1}{\pi \Omega_{N} m}
\int_{0}^{\infty}
\frac{d\nu}{\left[\nu^{2}+(\vec{\rm q}-\vec{\rm p})^{2}\right]}\cdot
\nonumber \\
&& \cdot
\left[
\stackrel{(N)}{D}\,^{(1)}_{\zeta^{"} \zeta}
(\nu;-ip,b^{2},-iq)^{\overline{\zeta}}+
(\vec{\sigma} \cdot \vec{\tau}) (\vec{\sigma} \cdot \vec{\rm v})^{\dagger}
\cdot
\right.
\nonumber \\
&& \cdot
\left.
\stackrel{(N)}{D}\,^{(2)}_{\zeta^{"} \zeta}
(\nu;-ip,b^{2},-iq)^{\overline{\zeta}}
\right]
+(subtractions);
\label{22}
\end{eqnarray}
where for the Born term:
\begin{eqnarray}
&&
\stackrel{(N)}{D}\,^{\stackrel{(1)}{(2)}}_{\zeta^{"} \zeta}
(\nu;-ip,b^{2},-iq)^{\overline{\zeta}}\Rightarrow
\Sigma^{(N)}(\nu)\
{\bf A}^{\stackrel{(1)}{(2)}}_{\zeta^{"} \zeta}(q,p)=\Sigma^{(N)}(\nu)\cdot
\nonumber \\
&& \cdot
\left\{
\begin{array}{c}
\zeta^{"} \zeta \\ 1
\end{array}
\right\}
\left[
(\varepsilon(q) \pm m \zeta^{"})
(\varepsilon(p) \pm m \zeta)
\right]^{1/2}=\Sigma^{(N)}(\nu)\
{\bf A}^{\stackrel{(1)}{(2)}}_{\zeta \zeta^{"}}(p,q)=
\nonumber \\
&&=\Sigma^{(N)}(\nu)\,\zeta
\left[
\frac{
\varepsilon(p)+m \zeta
     }
{\varepsilon(q)+m \zeta^{"}}
\right]^{1/2}
q
\left\{
\begin{array}{c}
\left(\eta^{\zeta^{"}}(q)\right)^{-1} \\ \eta^{\zeta}(p)
\end{array}
\right.
\label{23} \\
&&
\equiv
\Sigma^{(N)}(\nu)\,\zeta
\left[
\frac{
\varepsilon(p)+m \zeta
     }
{\varepsilon(q)+m \zeta^{"}}
\right]^{1/2}
q
\ M^{\stackrel{(1)}{(2)}}_{\zeta^{"} \zeta}(q,p);
\nonumber
\end{eqnarray}
\begin{equation}
\eta^{\zeta}(p) \equiv (w^{\zeta}(p)-m)/p \equiv
p/(w^{\zeta}(p)+m);\
\eta^{-\zeta}(p)=-(\eta^{\zeta}(p))^{-1}.
\label{24}
\end{equation}
It is convenient to pass to the quantities, depending from quantum
numbers $\zeta^{"},\zeta$ only via sheet's indices of the functions
$ w^{\zeta}(p),w^{\zeta^{"}}(q)$ i.e. via brunch indices
$\zeta=sgn\left(Re\ w(p)\right)$
of analytic functions $w(p)=\left(p^{2}+m^{2}\right)^{1/2}, w(q)$. This may
be achieved by putting in accordance with (\ref{23}) for $i=1,2$:
\begin{eqnarray}
& &
\stackrel{(N)}{D}\,^{(i)}_{\zeta^{"}\zeta}
(\nu;-ip,b^{2},-iq)^{\overline{\zeta}}=
\\
& &
=\zeta
\left[
\frac{\varepsilon(p)+m\zeta}{\varepsilon(q)+m\zeta^{"}}
\right]^{1/2}
q \ \stackrel{(N)}{\bf D}\,^{(i)}_{\zeta^{"}\zeta}
(\nu;-ip,b^{2},-iq)^{\overline{\zeta}}=
\nonumber \\
& &
=\zeta^{"}
\left[
\frac{\varepsilon(q)+m\zeta^{"}}{\varepsilon(p)+m\zeta}
\right]^{1/2}
p \ \stackrel{(N)}{\bf D}\,^{(i)}_{\zeta \zeta^{"}}
(\nu;-iq,b^{2},-ip)^{\overline{\zeta}};
\nonumber
\label{26}
\end{eqnarray}
then the symmetry (\ref{21}) takes the form: $i=1,2$
\begin{equation}
\stackrel{(N)}{\bf D}\,^{(i)}_{\zeta \zeta^{"}}
(\nu;-iq,b^{2},-ip)^{\overline{\zeta}}=
\frac{\eta^{\zeta^{"}}(q)}{\eta^{\zeta}(p)}
\stackrel{(N)}{\bf D}\,^{(i)}_{\zeta^{"} \zeta}
(\nu;-ip,b^{2},-iq)^{\overline{\zeta}}
\label{27}
\end{equation}

Now following to Fubini and Stroffolini \cite{FS} and to \cite{Kpr,Ktmf}
we calculate discontinuity over momentum transfer
$t=-(\vec{\rm q}-\vec{\rm p})^{2}$
from both sides of LS equation (\ref{20}) with the help of the
arbitrary dimension's relation:
\begin{eqnarray}
&&
\int \frac
{d\Omega_{N}(\vec{\rm n})\ \Xi^{[M]}_{l}(\vec{\rm n})}
{\left[X-(\vec{\tau} \cdot \vec{\rm n})
\right]
\left[Y-(\vec{\rm v} \cdot \vec{\rm n})
\right]}
=\frac{(-1)^{l}4\pi^{\lambda+1}}{2^{l}\Gamma(l+\lambda)}
\ \Xi^{[M]}_{l}
\left(
\vec{\tau}\frac{\partial}{\partial X}+
\vec{\rm v}\frac{\partial}{\partial Y}
\right)\cdot
\nonumber \\
&&
\cdot\int^{\infty}_{Z_{+}(X,Y)}
\frac{dZ}
{
\left[W(X,Y,Z)\right]^{1/2}
\left[Z-(\vec{\tau} \cdot \vec{\rm v}) \right]
}
\left(
\frac{W(X,Y,Z)}{Z^{2}-1}
\right)^{l-a_{N}};
\label{28}
\end{eqnarray}
\begin{eqnarray}
&&
W(X,Y,Z)=X^{2}+Y^{2}+Z^{2}-2XYZ-1;
\nonumber \\
&& Z_{\pm}(X,Y)=XY\pm
\left[
(X^{2}-1)(Y^{2}-1)
\right]^{1/2};
\nonumber
\end{eqnarray}
for spherical function $ \Xi^{[M]}_{l}(\vec{\rm n})$ on group SO(N) \cite{Vln},
and came to the following system of equation for T-matrix spectral density
over momentum transfer for operator (\ref{8}a):
\begin{eqnarray}
&&\stackrel{(N)}{\bf D}\,^{\stackrel{(1)}{(2)}}_{\zeta^{"} \zeta}
(\nu;-ip,b^{2},-iq)^{\overline{\zeta}}-
\Sigma^{(N)}(\nu)\
M^{\stackrel{(1)}{(2)}}_{\zeta^{"} \zeta}(q,p)=
\label{29}
\\
&&=\frac{(N-2)}{2m\pi} \left[
\frac{\Delta(q^{2},p^{2},-\nu^{2})}{\nu^{2}}
\right]^{a}
\int_{0}^{\nu} d\gamma\Sigma^{(N)}(\gamma)
\int_{0}^{\nu - \gamma} d\mu
\int_{\omega_{-}(\nu;\mu,\gamma;q,p)}^{\omega^{+}(\nu;\mu,\gamma;q,p)}
dk^{2}\cdot
\nonumber \\
&&\cdot
\left[
(\omega^{+}-k^{2})(k^{2}-\omega_{-})\right]^{-\frac{1}{2}-a} \ k\
\sum_{\zeta^{'}=\pm 1}
{\bf g}^{\zeta^{'}}(-ik;b)^{\overline{\zeta}}\cdot
\nonumber \\
&&
\cdot\left\{
\left[
M^{\stackrel{(1)}{(2)}}_{\zeta^{"} \zeta^{'}}(q,k)
+M^{\stackrel{(2)}{(1)}}_{\zeta^{"} \zeta^{'}}(q,k)
\left(
\frac{Z_{\nu}Y_{\mu}-X_{\gamma}}{Z^{2}_{\nu}-1}
\right)
\right]
\stackrel{(N)}{\bf D}\,^{\stackrel{(1)}{(2)}}_{\zeta^{'}\zeta}
(\mu;-ip,b^{2},-ik)^{\overline{\zeta}}
\right.
\nonumber \\
&&\left.
+M^{\stackrel{(1)}{(2)}}_{\zeta^{"} \zeta^{'}}(q,k)
\left(
\frac{Z_{\nu}X_{\gamma}-Y_{\mu}}{Z^{2}_{\nu}-1}
\right)
\stackrel{(N)}{\bf D}\,^{\stackrel{(2)}{(1)}}_{\zeta^{'} \zeta}
(\mu;-ip,b^{2},-ik)^{\overline{\zeta}}
\right\}
\nonumber
\end{eqnarray}
which is independent from subtractions in (\ref{22}).
Here we put $\mu_{0}=0$ for simplification of the formulas and the
following notations are accepted hereafter:
\[
{\bf g}^{\zeta^{'}}(-ik;b)^{\overline{\zeta}}=
\frac{(-1)}{(k^{2}+b^{2})}\frac{1}{2}
\left(
1+\frac{W^{\overline{\zeta}}(ib)}{w^{\zeta^{'}}(k)}
\right)
=
\left[
2w^{\zeta^{'}}(k)
\left(
W^{\overline{\zeta}}(ib)-w^{\zeta^{'}}(k)
\right)
\right]^{-1};
\]
\[
X_{\gamma}=\frac{q^{2}+k^{2}+\gamma^{2}}{2qk};\
Y_{\mu}=\frac{p^{2}+k^{2}+\mu^{2}}{2pk};\
Z_{\nu}\equiv Z(qp|\nu)=\frac{q^{2}+k^{2}+\nu^{2}}{2qp};
\]
\begin{eqnarray}
&&
2\nu^{2}\omega^{+}_{-}(\nu;\mu,\gamma;q,p)=
\nu^{2}(\nu^{2}-\mu^{2}-\gamma^{2})+q^{2}(\nu^{2}+\mu^{2}-\gamma^{2})+
\nonumber \\
&&
+p^{2}(\nu^{2}-\mu^{2}+\gamma^{2})\pm
\left[
\Delta(\nu^{2},\mu^{2},\gamma^{2}) \ \Delta(q^{2},p^{2},-\nu^{2})
\right]^{1/2};
\nonumber \\
&&
\Delta(a,b,c)=(a+b-c)^{2}-4ab;
\label{30}
\end{eqnarray}
For case (\ref{8}b) we must change $M^{(2)}\Rightarrow -M^{(2)}$.
For the same function $V(r)$ with different dimensions of $r_{(N)}$ and
$r_{(D)}$ its solutions are connected by simple  Weyl's integral
transformation: $i=1,2$
\begin{equation}
\stackrel{(D)}{\bf D}\,^{(i)}(\nu;\cdot\cdot)=
\frac{\Gamma(N/2)}{\Gamma(D/2)} 2\nu
\left(
\frac{d}{d \nu^{2}}
\right)^{n}
\int_{\mu_{0}}^{\nu}d \gamma
\frac{
(\nu^{2}-\gamma^{2})^{a_{D}-a_{N}+n-1}
     }
{\Gamma(a_{D}-a_{N}+n)}
\stackrel{(N)}{\bf D}\,^{(i)}(\gamma;\cdot\cdot),
\label{31}
\end{equation}
where integer number $n$ is restricted only by convergence
condition of this integral: $n\geq max [(D-N)/2;0]$.
This transformation is identical with Schrodinger case, and may
be checked by the same way \cite{Ktmf}.

For the Hamiltonian (\ref{18}) the formulas may be simplified by
choosing helicity representation for the spinors $ {\bf w}(\vec{\rm n}):\
(\vec{\sigma} \cdot \vec{\rm n})|{\bf w}_{\lambda}(\vec{\rm n})>=2\lambda
|{\bf w}_{\lambda}(\vec{\rm n})>$. Then, instead (\ref{23}), we have:
\begin{equation}
<\vec{\rm q},\mu |{\bf U}|\vec{\rm p},\lambda >=
\stackrel{*}{\cal D}\,^{(1/2)}_{\lambda \mu}({\bf R}_{\vec{\tau}\vec{\rm v}})
\left[
1-\frac{q^{2}+p^{2}-2\mu 2\lambda 2qp}{2 (2m)^{2}}
\right]
\frac{1}{2m}<\vec{\rm q}|V|\vec{\rm p}>;
\label{33}
\end{equation}
where potential $V(r)$ (\ref{10}) for N=3 also is assumed to be regular
(without subtractions).
Separating in (\ref{20}),(\ref{22}) spin-rotation matrix
\[
\stackrel{*}{\cal D}\,^{(1/2)}_{\lambda \mu}\left({\bf R}_
{\vec{\tau}\vec{\rm v}})= <{\bf w}_{\mu}(\vec{\tau})|
{\bf w}_{\lambda}(\vec{\rm v}\right)>,
\]
one can decompose the spectral density matrix onto the sum of two
orthogonal projectors with coefficients $D^{(1),(2)} \cite{Kpr}$:
\begin{equation}
D_{\mu \lambda}(\cdots)=D^{(1)}(\cdots)+2\mu 2\lambda \ D^{(2)}(\cdots);
\label{34}
\end{equation}
\[
2(\Pi^{+}_{1})_{\mu \lambda}=
(I+\sigma_{1})_{\mu \lambda}=1,
(for\  all\  \mu,\lambda);\
2(\Pi^{-}_{1})_{\mu \lambda}=(I-\sigma_{1})_{\mu \lambda}=2\mu 2\lambda;
\]
Then discontinuity calculation like above give the same system like
(\ref{29}) for N=3 with changed normalization
\[
\frac{q}{2m}
\stackrel{(N)}{\bf D}\,^{(i)}_{\zeta^{"} \zeta}
(\nu;-ip,b^{2},-iq)^{\overline{\zeta}}
\Rightarrow D^{(i)}(\cdots);\quad
\frac{q}{2m}M^{(i)}_{\zeta^{"}\zeta}(q,p)\rightarrow A^{(i)}(q,p);
\]
and with substitutions:
\be
\sum_{\zeta^{'}= \pm 1}{\bf g}^{\zeta^{'}}(-ik;b)^{\overline{\zeta}}
\rightarrow (k^{2}+b^{2})^{-1};
\label{35}
\ee
\begin{equation}
A^{(1)}(q,p)=1-\frac{1}{2}(q^{2}+p^{2})(2m)^{-2};\quad
A^{(2)}(q,p)=qp(2m)^{-2}.
\label{35*}
\end{equation}
Spectral density's equation for Hamiltonian (\ref{19}) has more simple
form \cite{Ktmf} with substitution:
\be
\Sigma^{(3)} (\nu)\Rightarrow
\Sigma_{1}(\nu) + \Sigma_{2}(\nu)(q^{2}+p^{2})(2m)^{-2}
\label{35**}
\ee
and in particular case $V_{1}=V_{2}$ appears from last system,
if one put on them $D^{(2)}=A^{(2)}=0$ eliminating all dependence from spin.

Let now shortly consider analytic properties of the spectral density
over energetic variables $q,p$. It may be shown \cite{Kds}, that due to
volterrian property of eq.(\ref{29}) providing convergence of its iteration
serie, the spectral density possess analytic continuation to the domain
\cite{Ktmf}
\begin{eqnarray}
& &
\left.
\begin{array}{c}
p=i\varrho,\  q=iu,\  k=i\alpha;\  \varrho >0;\  0<\nu<u-\varrho;
\\
\left[\Delta (q^{2},p^{2},-\nu^{2})\right]^{1/2}
=\ e^{i\pi}\
\left[\Delta (u^{2},\varrho^{2},\nu^{2})\right]^{1/2};
\\
\omega^{+}_{-}(\nu;\mu,\gamma;q,p)=\ e^{i\pi}\
\Lambda^{+}_{-}(\nu;\mu,\gamma;u,\varrho);
\end{array}
\right\}
\label{36}
\end{eqnarray}
where we put:
\begin{eqnarray}
&&
\Lambda^{+}_{-}(\nu;\mu,\gamma;u,\varrho)=
\Lambda^{0}(\nu;\mu,\gamma;u,\varrho)\pm\frac{1}{2\nu^{2}}
\left[
\Delta(\nu^{2},\mu^{2},\gamma^{2})\ \Delta(u^{2},\varrho^{2},\nu^{2})
\right]^{1/2};
\nonumber
\end{eqnarray}
\begin{eqnarray}
&&
2\nu^{2}\Lambda^{0}(\nu;\mu,\gamma;u,\varrho)=
\nu^{2}(\mu^{2}+\gamma^{2}-\nu^{2})
+u^{2}(\nu^{2}+\mu^{2}-\gamma^{2})+
\nonumber \\
&&
+\varrho^{2}(\nu^{2}-\mu^{2}+\gamma^{2}),
\label{37}
\end{eqnarray}
and that continued functions satisfy to system (\ref{29}) continued
to this domain (see bellow).

\section{\rm Generalizations of the off-shell Jost function method
for Dirac operator}

Let us now turn to generalizations of the off-shell Jost
functions method. Such preliminaries is necessary to establish
the relation in question between Jost function and T-matrix momentum
transfer spectral density. There are two ways to introduce off-shell
Jost functions (OSJF). The first one derive it only for local
potential from solution of nonhomogeneous radial Schrodinger (or Dirac)
equations i.e. off-shell Jost solution (OSJS). The second one relate
OSJF with half-off-shell partial amplitude. Both this ways are
equivalent obviously for local nonsingular potentials,
successfully added each other for singular and nonlocal potentials.
Although the ideas of this method is not new \cite{FW,Psq},
we outline here its main points in modified form, convenient for our
aims to get its generalization on complex value of total angular momentum
$J_{\rm N}$ and demonstrate its applicability for a wide class of
operators.

We begin with the second way \cite{Ktmf} introducing off-shell partial
amplitudes by expansion of T-matrix (\ref{22}):
\begin{eqnarray}
&&
{\bf (}\vec{\rm q},\zeta^{"}|
\stackrel{(N)}{\bf T}
(W^{\overline{\zeta}}(ib))|
\vec{\rm p},\zeta{\bf )}=
-\frac{2(qp)^{a}}{\pi q}
\zeta^{"}\left[\frac{\varepsilon(q)+m\zeta^{"}}{\varepsilon(p)+m\zeta}
\right]^{1/2}\cdot
\nonumber \\
&& \cdot
\sum^{\infty}_{J=\lambda_{N}}
\sum_{\xi=\pm 1}
\Pi_{\kappa_{\xi}}(\vec{\tau},\vec{\rm v})\quad
T^{\zeta^{"}\zeta}_{\kappa_{\xi}}(q,p;b^{2})^{\overline{\zeta}}.
\label{38}
\end{eqnarray}
For sufficiently large value of $J_{\rm N}$ it possess a Froissart-Gribov
integral representation:
\begin{eqnarray}
& &
T^{\zeta^{"} \zeta}_{\kappa_{\xi}}(q,p;b^{2})^{\overline{\zeta}}=
-\frac{4 \pi e^{-i \pi a}}{4m \Omega_{N} \pi^{a}}
\int_{\mu_{0}}^{\infty} d \nu
\left[
Q^{a}_{L_{\xi}}(Z_{\nu}) \stackrel{(N)}{\bf D}\,^{(1)}_{\zeta \zeta^{"}}
(\nu;-iq,b^{2},-ip)^{\overline{\zeta}}+
\right.
\nonumber \\
& &
\left.+
Q^{a}_{L_{-\xi}}(Z_{\nu})
\stackrel{(N)}{\bf D}\,^{(2)}_{\zeta \zeta^{"}}
(\nu;-iq,b^{2},-ip)^{\overline{\zeta}}
\right]
\left[
\Delta(q^{2},p^{2},-\nu^{2})
\right]^{-a/2},
\label{39}
\end{eqnarray}
($Z_{\nu}$ is defined in (\ref{30}))
which for Born term (see (\ref{23})) take place without restriction.
The OSJF $ F^{\zeta}_{\kappa_{\xi}}(\varrho,-ik)^{\overline{\zeta}} $
is introduced as two variable's function analytic in the domain
\be
\Bigl(
\varrho,\zeta;k,\overline{\zeta}:\, Re\,\varrho > 0,\,\varrho \not\in [m,+\infty),
\,\zeta =\pm 1;\
|Im\,k|<\mu_{0},\,\pm ik \not\in  [m,+\infty),\,\overline{\zeta}=\pm 1
\Bigr)
\label{*}
\ee
which decompose the partial half-off-shell amplitude
\[
\left.T^{\zeta^{"}{\overline{\zeta}}}_{\kappa_{\xi}}(q,k;b^{2})^
{\overline{\zeta}}
\right|_{b=0\mp ik}=
T^{\zeta^{"}\overline{\zeta}(\pm)}_{\kappa_{\xi}}(q,k)
\]
according to \cite{FW}:
\begin{equation}
T^{\zeta^{"} \overline{\zeta} (\pm)}_{\kappa_{\xi}}(q,k)=
\left(
\frac{k}{q}
\right)^{L_{\xi}}
\left[
F^{\zeta^{"}}_{\kappa_{\xi}}(iq,-ik)^{\overline{\zeta}}-
F^{\zeta^{"}}_{\kappa_{\xi}}(-iq,-ik)^{\overline{\zeta}}
\right]
\left[
2iF^{\overline{\zeta}}_{\kappa_{\xi}}(\mp ik)
\right]^{-1}.
\label{40}
\end{equation}
It means, that Jost function is simply related with OSJF:
\be
F^{\overline{\zeta}}_{\kappa_{\xi}}(\mp ik,-ik)^{\overline{\zeta}}=
F^{\overline{\zeta}}_{\kappa_{\xi}}(\mp ik).
\label{41}
\ee
However, inversion of the decomposition (\ref{40}) now is not as
straightforward as for Schrodinger case \cite{Ktmf}. One can see, that
all mentioned above properties of OSJF hold for the following ansatz:
\begin{eqnarray}
& &
F^{\zeta}_{\kappa_{\xi}}(\varrho,-ik)^{\overline{\zeta}}-
Z^{\zeta}_{\kappa_{\xi}}(\varrho^{2},k^{2})^{\overline{\zeta}}=
F^{\overline{\zeta}}_{\kappa_{\xi}}(\mp ik)
\frac{1}{\pi}
\int_{0}^{\infty} ds^{2}
\left(
\frac{s}{k}
\right)^{L_{\xi}}
\cdot
\nonumber \\
& &
\cdot
\sum_{\zeta^{'}=\pm 1} {\bf g}^{\zeta^{'}} (-is;\varrho)^{\zeta}\
N^{\zeta}_{\xi}(\varrho,-is)^{\zeta^{'}}
T^{\zeta^{'}\overline{\zeta} (\pm)}_{\kappa_{\xi}}(s,k);
\label{42}
\end{eqnarray}
where we introduce the notation:
\begin{equation}
N^{\zeta}_{\xi}(\varrho,b)^{\overline{\zeta}}=
\left[
\frac{
W^{\overline{\zeta}}(ib)+m
      }
{
w^{\zeta}(i\varrho)+m
}
\right]^{\frac{1-\xi}{2}}
=
\left[
\frac{
\eta^{\zeta}(i\varrho) \ b
      }
{
\eta^{\overline{\zeta}}(ib) \ \varrho
}
\right]^{\frac{1-\xi}{2}};\
N^{\zeta}_{\xi}(\varrho,\varrho)^{\zeta}=1.
\label{43}
\end{equation}
and meromorfic on two-sheet's Remanian surface (\ref{*}) unknown function
of $\varrho^{2}$, disappearing in difference (\ref{40}) and satisfying
to condition:
\begin{equation}
\left.
Z^{\overline{\zeta}}_{\kappa_{\xi}}(-k^{2},k^{2})^{\overline{\zeta}}
\right|_{loc}=1.
\label{44}
\end{equation}
Due to this condition the asatz (\ref{42}) for $\varrho=\mp ik$ convert
in accordance with (\ref{41}) to general representation for Jost function.
The last follows directly from abstract definition (\ref{1}) with the help
of known reasoning \cite{BrFLS}, using decomposition of partial
Green function (\ref{3}) into Volterrian and separable parts (see Appendix),
and the relation for physical solution of radial Dirac equation (see bellow)
which reads: $(\kappa=\kappa_{\xi}),$
\begin{equation}
-\frac{V(r)}{\eta^{\zeta}(k)}\ \psi^{(\pm) \zeta}_{\kappa V}(k,r)=
\frac{1}{\pi}
\int_{0}^{\infty} ds^{2} \sum_{\zeta^{'}=\pm 1}
\frac{\phi^{\zeta^{'}}_{\kappa 0}(s,r)}{2w^{\zeta^{'}}(s)\,\eta^{\zeta^{'}}(s)}
\ T^{\zeta^{"}\zeta(\pm)}_{\kappa}(s,k).
\label{45}
\end{equation}
Substituting the partial LS-equation (which is a Fredholm-type equation)
\begin{eqnarray}
& &
T^{\zeta^{"}\overline{\zeta} (\pm)}_{\kappa}(q,k)-
T^{\zeta^{"}\overline{\zeta}}_{\kappa 0}(q,k)=
-\frac{1}{\pi}
\int_{0}^{\infty} ds^{2}
\sum_{\zeta^{'}=\pm 1} {\bf g}^{\zeta^{'}} (-is;\mp ik)^{\overline{\zeta}}
\cdot
\nonumber \\
& &
\cdot
T^{\zeta^{"}\zeta^{'}}_{\kappa 0}(q,s)\
T^{\zeta^{'}\overline{\zeta} (\pm)}_{\kappa}(s,k),
\label{46}
\end{eqnarray}
to the right hand side of ansatz (\ref{42}), and using its particular form
for $\varrho=\mp ik$ (clf.(\ref{41})) in the first of appearing items,
one has for this r.h.s. the expression:
\[
Z^{\overline{\zeta}}_{\kappa_{\xi}}(-k^{2},k^{2})^{\overline{\zeta}}\,
H^{\zeta \overline{\zeta}}_{\kappa_{\xi}}(\varrho,k)-
F^{\overline{\zeta}}_{\kappa_{\xi}}(\mp ik)
\frac{1}{\pi}
\int_{0}^{\infty} ds^{2}
\left(
\frac{s}{k}
\right)^{L_{\xi}} \sum_{\zeta^{'}=\pm 1}
{\bf g}^{\zeta^{'}} (-is;\mp ik)^{\overline{\zeta}}
\cdot
\]
\be
\cdot
T^{\zeta^{'} \overline{\zeta}
(\pm)}_{\kappa_{\xi}}(s,k) \left[ H^{\zeta \zeta^{'}}_{\kappa_{\xi}}(\varrho,s)-
N^{\overline{\zeta}}_{\xi}(-ik,-is)^{\zeta^{'}}\,
H^{\zeta \overline{\zeta}}_{\kappa_{\xi}}(\varrho,k)
\right];
\label{47}
\ee
where the auxiliary kernel is introduced:
\begin{equation}
H^{\zeta \overline{\zeta}}_{\kappa_{\xi}}(\varrho,k)=
\frac{1}{\pi}
\int_{0}^{\infty} ds^{2}
\left(
\frac{s}{k}
\right)^{L_{\xi}}
\sum_{\zeta^{'}=\pm 1} {\bf g}^{\zeta^{'}} (-is;\varrho)^{\zeta}\,
N^{\zeta}_{\xi}(\varrho,-is)^{\zeta^{'}}\,
T^{\zeta^{'} \overline{\zeta}}_{\kappa_{\xi} 0}(s,k).
\label{48}
\end{equation}
The relation which following \cite{Kds} from formula (\ref{65*}) (see bellow)
for \\
$
Re\ j>-1-a_{N};\ {\rm T}(u\varrho|\nu)=\left(u^{2}+\varrho^{2}-\nu^{2}\right)/2u\varrho
$:
\[
\int_{u+\nu}^{\infty}
\frac{
d\alpha \quad
P^{a}_{j} \left({\rm T}(u\alpha|\nu) \right)
      }
{
\left[
\Delta (u^{2},\alpha^{2},\nu^{2})
\right]^{a/2}
\left(
\alpha^{2}+k^{2}
\right)
}
\left(
\frac{u}{\alpha}
\right)^{j}=
\]
\be
=\int_{0}^{\infty}
\frac{
ds^{2} \quad
Q^{a}_{j} \left( Z(sk|\nu) \right) e^{-i\pi a}
      }
{
2\pi k
\left[
\Delta (s^{2},k^{2},-\nu^{2})
\right]^{a/2}
\left(
s^{2}+u^{2}
\right)
}
\left(
\frac{s}{k}
\right)^{j};
\label{49*}
\ee
and easily verifying formulae
\[
\sum_{\zeta^{'}=\pm 1}
\frac{1}{2}
\left(
1+\frac{W^{\zeta}(k)}{w^{\zeta^{'}}(s)}
\right)
\left(
\frac{w^{\zeta^{'}}(s)\pm m}{W^{\zeta}(k)\pm m}
\right)^{\frac{1}{2}(1-\xi)}=1;
\]
\be
\sum_{\zeta^{'}=\pm 1}
\frac{1}{2}
\left(
1+\frac{W^{\zeta}(k)}{w^{\zeta^{'}}(s)}
\right)
\left(
\frac{w^{\zeta^{'}}(s)\pm m}{W^{\zeta}(k)\pm m}
\right)^{\frac{1}{2}(1-\xi)}
\left[
\eta^{\zeta^{'}}(s)
\right]^{\pm 1}
=\left(
\frac{k}{s}
\right)^{\xi}
\left[
\eta^{\zeta}(k)
\right]^{\pm 1};
\label{49**}
\ee
allow to rewrite the auxiliary kernel (\ref{48}) as:
\begin{equation}
H^{\zeta \overline{\zeta}}_{\kappa_{\xi}}(\varrho,k)=
\int_{\varrho+\mu_{0}}^{\infty} d\alpha
\left(
\frac{\varrho}{\alpha}
\right)^{L_{\xi}}
\sum_{\zeta^{'}=\pm 1}
{\bf g}^{\zeta^{'}} (\alpha;-ik)^{\overline{\zeta}}\,
N^{\zeta^{'}}_{\xi}(\alpha,-ik)^{\overline{\zeta}}\,
{\bf K}^{\zeta^{'} \zeta}_{\kappa_{\xi}}(\alpha,\varrho);
\label{49}
\end{equation}
where the new Volterrian kernels are introduced for the case (\ref{8}a):
\begin{equation}
K_{j}(u,\varrho)=\frac{4\pi}{\Omega_{N}\pi^{a}}
\int_{\mu_{0}}^{u-\varrho} d\nu
P^{a}_{j} \left({\rm T}(u\varrho |\nu) \right)
\frac{\Sigma^{(N)}(\nu)}
{
\left[ \Delta (u^{2},\varrho^{2},\nu^{2}) \right]^{a/2}
};
\label{50*}
\end{equation}
\begin{equation}
{\bf K}^{\zeta^{'}\zeta}_{\kappa_{\xi}}(u,\varrho)=
\frac{iu}{2m}
\left[
\frac{1}{\eta^{\zeta^{'}}(iu)} \ K_{L_{\xi}}(u,\varrho)
+\eta^{\zeta}(i\varrho) \ K_{L_{-\xi}}(u,\varrho)
\right].
\label{50}
\end{equation}
Here for the case (\ref{8}b) the second term has opposite sign
(clf. remark after (\ref{30})) and the branch $w^{\zeta}(p)$ takes
value at $p=i\varrho+0, \varrho>m:\
w^{\zeta}(i\varrho) \Rightarrow i\zeta \sqrt{\varrho^{2}-m^{2}} $.
This choice is conventional and does not affect on sum over the
sheets $\zeta=\pm 1$, for which kinematical cuts $\pm \varrho>m$ disappears.
Substitution of the (\ref{49}) and repeating use of ansatz (\ref{42})
under the $\alpha$-integral, converts the relations (\ref{42}),
(\ref{47}) to the following Volterra-type equation for OSJF:
\begin{eqnarray}
& &
F^{\zeta}_{\kappa_{\xi}}(\varrho,-ik)^{\overline{\zeta}}
-Z^{\zeta}_{\kappa_{\xi}}(\varrho^{2},k^{2})^{\overline{\zeta}}=
\int_{\varrho+\mu_{0}}^{\infty} d\alpha
\left(
\frac{\varrho}{\alpha}
\right)^{L_{\xi}}
\sum_{\zeta^{'}=\pm 1} {\bf g}^{\zeta^{'}} (\alpha;-ik)^{\overline{\zeta}}
\cdot
\nonumber \\
& &
\cdot
{\bf K}^{\zeta^{'}\zeta}_{\kappa_{\xi}}(\alpha,\varrho)
\left[
F^{\zeta}_{\kappa_{\xi}}(\alpha,-ik)^{\overline{\zeta}}
-Z^{\zeta^{'}}_{\kappa_{\xi}}(\alpha^{2},k^{2})^{\overline{\zeta}}
+N^{\zeta^{'}}_{\xi}(\alpha,-ik)^{\overline{\zeta}}
\right].
\label{51}
\end{eqnarray}
A natural choice of OSJF's normalization now is given by the relation:
\begin{equation}
\left.
Z^{\zeta}_{\kappa_{\xi}}(\varrho^{2},k^{2})^{\overline{\zeta}}
\right|_{lok.nonsin.}=
N^{\zeta}_{\xi}(\varrho,-ik)^{\overline{\zeta}},
\label{52}
\end{equation}
where function in the right hand side obviously satisfy to all
conditions (\ref{*}),(\ref{44}) written out for the left one. It
transforms the equation (\ref{51}) for $b=-ik$ to the following form:
\begin{eqnarray}
& &
F^{\zeta}_{\kappa_{\xi}}(\varrho,b)^{\overline{\zeta}}-
N^{\zeta}_{\xi}(\varrho,b)^{\overline{\zeta}}=
\int_{\varrho+\mu_{0}}^{\infty} du
\left(
\frac{\varrho}{u}
\right)^{L_{\xi}}
\sum_{\zeta^{'}=\pm 1} {\bf g}^{\zeta^{'}} (u;b)^{\overline{\zeta}} \cdot
\nonumber \\
& &
\cdot
\left\{
\begin{array}{c}
F^{\zeta^{'}}_{\kappa_{\xi}}(u,b)^{\overline{\zeta}}\ \
{\bf K}^{\zeta^{'}\zeta}_{\kappa_{\xi}}(u,\varrho).
\\
N^{\zeta^{'}}_{\xi}(u,b)^{\overline{\zeta}}\ \
{\bf a}^{\zeta^{'}\zeta}_{\kappa_{\xi}}(u,\varrho;b^{2})^{\overline{\zeta}}.
\end{array}
\right.
\label{53}
\end{eqnarray}
Here the second line expresses solution of the first line's equation
via Volterrian resolvent kernel $
{\bf a}^{\zeta^{'}\zeta}_{\kappa_{\xi}}(u,\varrho;b^{2})^{\overline{\zeta}}$
which therefor satisfy in its turn to Volterra equations:
\begin{eqnarray}
& &
{\bf a}^{\zeta^{"}\zeta}_{\kappa}(u,\varrho;b^{2})^{\overline{\zeta}}-
{\bf K}^{\zeta^{"}\zeta}_{\kappa}(u,\varrho)=
\int_{\varrho+\mu_{0}}^{u-\mu_{0}} d\alpha
\sum_{\zeta^{'}=\pm 1} {\bf g}^{\zeta^{'}} (\alpha;b)^{\overline{\zeta}}
\cdot
\nonumber \\
& &
\cdot
\left\{
\begin{array}{c}
{\bf K}^{\zeta^{"}\zeta^{'}}_{\kappa}(u,\alpha)\ \
{\bf a}^{\zeta^{'}\zeta}_{\kappa}(\alpha,\varrho;b^{2})^{\overline{\zeta}};
\\
{\bf a}^{\zeta^{"}\zeta^{'}}_{\kappa}(u,\alpha;b^{2})^{\overline{\zeta}}
\ \ {\bf K}^{\zeta^{'}\zeta}_{\kappa}(\alpha,\varrho);
\end{array}
\right.
\label{54}
\end{eqnarray}
compatible with the following symmetry properties (cmp.(\ref{27})):
\begin{equation}
{\bf a}^{\zeta^{'}\zeta}_{\kappa}(u,\varrho;b^{2})^{\overline{\zeta}}\ =
\ \frac
{
\eta^{\zeta}(i\varrho)\ u
  }
{
\eta^{\zeta^{'}}(iu)\ \varrho
}\
{\bf a}^{\zeta\zeta^{'}}_{\kappa}(-\varrho,-u;b^{2})^{\overline{\zeta}}.
\label{55}
\end{equation}

Now in the equation (\ref{54}) it is possible to make an exact factorization
of dependency from $J_{\rm N}$ which is the main observation of this work.
Its form is prompted by the expression for the kernel (\ref{50}) and reads:
\begin{eqnarray}
& &
{\bf a}^{\zeta^{"} \zeta}_{\kappa_{\xi}}(u,\varrho;b^{2})^{\overline{\zeta}}=
\frac{4 \pi}{\Omega_{N} \pi^{a}} \frac{iu}{2m}
\int_{\mu_{0}}^{u-\varrho} d \nu
\left[
P^{a}_{L_{\xi}} \left({\rm T}(u\varrho | \nu )\right)
\stackrel{(N)}{\tilde{\bf D}}\,^{(1)}_{\zeta^{"}\zeta }
(\nu;\varrho,b^{2},u)^{\overline{\zeta}}+
\right.
\nonumber \\
& &
\left.
+P^{a}_{L_{-\xi}} \left({\rm T}(u\varrho | \nu )\right)
\stackrel{(N)}{\tilde{\bf D}}\,^{(2)}_{\zeta^{"}\zeta }
(\nu;\varrho,b^{2},u)^{\overline{\zeta}}
\right]
\left[
\Delta (u^{2},\varrho^{2},\nu^{2})
\right]^{-a/2}.
\label{56}
\end{eqnarray}
It may be checked by induction with the help of relations for arbitrary
integrable function $\cal H(\alpha)$, for arbitrary complex $l$, and
$Re\  a<1/2$ \cite{Ktmf,Kds}:
\begin{eqnarray}
&&
\int_{\varrho + \mu}^{u-\gamma} d\alpha\
\frac{
P^{a}_{l} \left({\rm X}(u\alpha |\gamma ) \right)
      }
{
\left[ \Delta (u^{2},\alpha^{2},\gamma^{2}) \right]^{a/2}
}\
\frac{
P^{a}_{l} \left({\rm Y}(\alpha \varrho|\mu) \right)
      }
{
\left[ \Delta (\alpha^{2},\varrho^{2},\mu^{2}) \right]^{a/2}
}\
{\cal H}(\alpha)=
\nonumber \\
&&
=\int_{\gamma + \mu}^{u-\varrho} d \nu
P^{a}_{l} \left({\rm T}(u \varrho|\nu ) \right)
\frac{\left[ \Delta (u^{2},\varrho^{2},\nu^{2}) \right]^{a/2}}
{(\nu^{2})^{a}\Gamma(\lambda)\sqrt{\pi}}
\int_{\Lambda_{-}(\nu;\mu,\gamma;u,\varrho)}
^{\Lambda^{+}(\nu;\mu,\gamma;u,\varrho)} d\alpha^{2}
\ {\cal H}(\alpha)\cdot
\nonumber \\
&&
\cdot
\left(
(\Lambda^{+}-\alpha^{2})(\alpha^{2}-\Lambda_{-})
\right)^{-a-1/2};
\label{57}
\end{eqnarray}
and its analogy for the products
$P^{a}_{l}({\rm X}_{\gamma}) \cdot P^{a}_{l \pm 1}({\rm Y}_{\mu})$, following
by differentiation from multiplication formula and recurrence relations
for Legendre
functions $P^{a}_{l}({\rm T})$ \cite{Bt-Er,Vln}. With this relations,
substitution of(\ref{56}) into equations (\ref{54}) leads to the system
for independent from $J_{\rm N}$ and $L_{\xi} $ functions:
\begin{eqnarray}
&&
\stackrel{(N)}{\tilde{\bf D}}\,^{\stackrel{(1)}{(2)}}_{\zeta^{"}\zeta}
(\nu;\varrho,b^{2},u)^{\overline{\zeta}}-\Sigma^{(N)}(\nu)\,
M^{\stackrel{(1)}{(2)}}_{\zeta^{"}\zeta}
(iu,i\varrho)=
\label{58}
 \\
&&
=
\frac{(N-2)}{2m\pi}
\left[
\frac{ \Delta (u^{2},\varrho^{2},\nu^{2})}{\nu^{2}}
\right]^{a}
\int_{\mu_{0}}^{\nu -\mu_{0}} d\gamma \,\Sigma^{(N)}(\nu)
\int_{\mu_{0}}^{\nu -\gamma} d\mu
\int_{\Lambda_{-}(\nu;\mu,\gamma;u,\varrho)}
^{\Lambda^{+}(\nu;\mu,\gamma;u,\varrho)} d\alpha^{2}
\cdot
\nonumber \\
&&
\cdot
\left[
(\Lambda^{+}-\alpha^{2})(\alpha^{2}-\Lambda_{-})
\right]^{-\frac{1}{2}-a}
i\alpha
\sum_{\zeta^{'}=\pm 1}{\bf g}^{\zeta^{'}}(\alpha;b)^{\overline{\zeta}}
\cdot
\nonumber \\
&&
\cdot
\left\{
\left[
M^{\stackrel{(1)}{(2)}}_{\zeta^{"} \zeta^{'}}(iu,i\alpha)
+M^{\stackrel{(2)}{(1)}}_{\zeta^{"} \zeta^{'}}(iu,i\alpha)
\left(
\frac{{\rm T}_{\nu}{\rm Y}_{\mu}-{\rm X}_{\gamma}}{{\rm T}^{2}_{\nu}-1}
\right)
\right]
\stackrel{(N)}{\tilde{\bf D}}\,^{\stackrel{(1)}{(2)}}_{\zeta^{'} \zeta}
(\mu;\varrho,b^{2},\alpha)^{\overline{\zeta}}
\right.
\nonumber \\
&&
\left.
+M^{\stackrel{(1)}{(2)}}_{\zeta^{"} \zeta^{'}}(iu,i\alpha)
\left(
\frac{{\rm T}_{\nu}{\rm X}_{\gamma}-{\rm Y}_{\mu}}{{\rm T}^{2}_{\nu}-1}
\right)
\stackrel{(N)}{\tilde{\bf D}}\,^{\stackrel{(2)}{(1)}}_{\zeta^{'} \zeta}
(\mu;\varrho,b^{2},\alpha)^{\overline{\zeta}}
\right\}.
\nonumber
\end{eqnarray}
Here: ${\rm X}_{\gamma}={\rm X}(u \alpha|\gamma),{\rm Y}_{\mu}=
{\rm Y}(\varrho \alpha|\mu)$ are defined analogously to
${\rm T}_{\nu}={\rm T}(u \varrho|\nu)$ in (\ref{49*}), and
$\Lambda^{+}_{-}$ are given in (\ref{37}).
Comparing this system with the one in (\ref{29}), and keeping in mind
formulas of the analytic continuation (\ref{36}), it is not difficult
to see that the systems and theirs solutions are analytic continuations
of one another: when
$p=i\varrho;\ q=iu;\ k=i\alpha;\ \varrho>0;\ u-\varrho>\nu>0;\ Z(qp|\nu)=
{\rm T}(u\varrho|\nu)$, etc.,
then for $i=1,2$
\begin{equation}
\stackrel{(N)}{\bf D}\,^{(i)}_{\zeta^{"} \zeta}(\mu;-ip,b^{2},-iq)
^{\overline{\zeta}}\ =\
\stackrel{(N)}{\tilde{\bf D}}\,^{(i)}_{\zeta^{"} \zeta}(\mu;\varrho,b^{2},u)
^{\overline{\zeta}}.
\label{59}
\end{equation}
Eq. (\ref{53}),(\ref{56}) give the following representation for
the Jost function:
\begin{eqnarray}
&&
F^{\overline{\zeta}}_{\kappa_{\xi}}(b)\equiv
F^{\overline{\zeta}}_{\kappa_{\xi}}(b,b)^{\overline{\zeta}}=
\nonumber \\
&&
=1+\int_{b+\mu_{0}}^{\infty}du
\left(
\frac{b}{u}
\right)^{L_{\xi}}
\sum_{\zeta^{'}=\pm 1}{\bf g}^{\zeta^{'}}(u,b)^{\overline{\zeta}}\
N^{\zeta^{'}}_{\xi}(u,b)^{\overline{\zeta}}\
\Phi^{\zeta^{'}\overline{\zeta}}_{\kappa_{\xi}}(u,b);
\label{60}
\end{eqnarray}
where we put:
\be
\Phi^{\zeta^{"}\overline{\zeta}}_{\kappa_{\xi}}(u,b)=
{\bf a}^{\zeta^{"}{\overline{\zeta}}}_{\kappa_{\xi}}(u,b;b^{2})^
{\overline{\zeta}}.
\label{61}
\ee

The results obtained here may be confirmed independently via the first
of mentioned ways,
dealing with off-shell Jost solution (OSJS) which satisfy to radial
Dirac equation: (here $\sigma_{1,2,3}$ are usual Pauly matrices)
\[
\left({\cal L}^{\kappa\overline{\zeta}}_{r}(b)-{\bf V}\right)\
{\bf J}^{\zeta}_{\kappa}(\varrho,b;r)^{\overline{\zeta}}\ =\
\left[
W^{\overline{\zeta}}(ib)-w^{\zeta}(i\varrho)
\right]\
X^{\zeta}_{\kappa}(\varrho,r);
\]
\begin{equation}
{\cal L}^{\kappa\overline{\zeta}}_{r}(b)-{\bf V}=
(i\sigma_{2})\partial_{r}-
\sigma_{1}\kappa_{\xi} r^{-1}-
\sigma_{3}m+W^{\overline{\zeta}}(ib)-{\bf V}(r),
\label{62}
\end{equation}
with potential ${\bf V}(r)$ defined in (\ref{8}a,b), and
with boundary condition at $r\rightarrow \infty$:
\[
{\bf J}^{\zeta}_{\kappa}(\varrho,b;r)^{\overline{\zeta}}
\rightarrow X^{\zeta}_{\kappa}(\varrho,r)
\rightarrow e^{-\varrho r}
\left|
\begin{array}{c}
1 \\ i\eta^{\zeta}(i\varrho)
\end{array}
\right|,
\]
where $ X^{\zeta}_{\kappa}(\varrho,r)$ is corresponding free solution (see
Appendix).
One may check that from (\ref{62}),(\ref{54}) for OSJS the relations follow:
\begin{eqnarray}
&&
{\bf J}^{\zeta}_{\kappa}(\varrho,b;r)^{\overline{\zeta}}-
X^{\zeta^{'}}_{\kappa}(\varrho,r)=
\int_{\varrho+\mu_{0}}^{\infty} du
\sum_{\zeta^{'}=\pm 1} {\bf g}^{\zeta^{'}} (u;b)^{\overline{\zeta}}
\cdot
\label{64} \\
&&
\cdot
\left\{
\begin{array}{c}
{\bf K}^{\zeta^{'}\zeta}_{\kappa}(u,\varrho)\
{\bf J}^{\zeta^{'}}_{\kappa}(u,b;r)^{\overline{\zeta}}
\\
{\bf a}^{\zeta^{'}\zeta}_{\kappa}(u,\varrho;b^{2})^{\overline{\zeta}}\
X^{\zeta^{'}}_{\kappa}(u,r)
\end{array}
\right. ;
\nonumber
\end{eqnarray}
\begin{equation}
{\bf V}(r)\,{\bf J}^{\zeta}_{\kappa}(\varrho,b;r)^{\overline{\zeta}}=
\int_{\varrho+\mu_{0}}^{\infty} du
\sum_{\zeta^{'}=\pm 1}
\frac{1}{2w^{\zeta^{'}}(iu)}
X^{\zeta^{'}}_{\kappa}(u,r) \
{\bf a}^{\zeta^{'}\zeta}_{\kappa}(u,\varrho;b^{2})^{\overline{\zeta}}.
\label{65}
\end{equation}
The Born version of the last relation corresponding to substitutions:
\[
{\bf J}^{\zeta}_{\kappa}(\varrho,b;r)^{\overline{\zeta}}
\rightarrow X^{\zeta}_{\kappa}(\varrho,r);\
{\bf a}^{\zeta^{'}\zeta}_{\kappa}(u,\varrho;b^{2})^{\overline{\zeta}}\
\rightarrow {\bf K}^{\zeta^{'}\zeta}_{\kappa}(u,\varrho),
\]
follows directly \cite{Kds} from definitions (\ref{50*}),(\ref{50}) and
the formula:
\be
\int_{\varrho+\nu}^{\infty} du\,
\frac{P^{a}_{j} \left({\rm T}(u\varrho |\nu) \right)}
{
\left[ \Delta (u^{2},\varrho^{2},\nu^{2}) \right]^{a/2}
}\
\chi_{j}(ur)
=\frac{1}{r}
\left(
\frac{r}{2\nu}
\right)^{a} \chi_{-a}(\nu r)\
\chi_{j}(\varrho r).
\label{65*}
\ee

Eq. (\ref{53}) for OSJF arise now by convolution of OSJS (\ref{64}) with
independent from $J_{\rm N}$ spinor:
$ \left( \stackrel{\rm (T)}{\cal S}_{\xi}(\varrho,b)^{\overline{\zeta}}
\right)_{1;2}= \left[ \frac{1+\xi}{2};
\frac{1-\xi}{2}b\left(i\varrho\  \eta^{\overline{\zeta}}(ib) \right)^{-1}
\right];
$
\begin{equation}
F^{\zeta}_{\kappa_{\xi}}(\varrho,b)^{\overline{\zeta}}=
\lim_{r \rightarrow 0}
\frac{
\sqrt{\pi}
      }
{
\Gamma \left( |\kappa_{\xi}|+\frac{1}{2} \right)
}
\left(
\frac{\varrho r}{2}
\right)^{|\kappa_{\xi}|}
\left\{
\stackrel{\rm (T)}{\cal S}_{\xi}(\varrho,b)^{\overline{\zeta}}\
{\bf J}^{\zeta}_{\kappa_{\xi}}(\varrho,b;r)^{\overline{\zeta}}
\right\}.
\label{66}
\end{equation}

We end this section by observation of the simple consequence of CPT-symmetry
for the Jost function:
\be
a) F^{\overline{\zeta}}_{\kappa}(b|{\it g})=
F^{-\overline{\zeta}}_{-\kappa}(b|-{\it g});
\qquad
b) F^{\overline{\zeta}}_{\kappa}(b|{\it g})=
F^{-\overline{\zeta}}_{-\kappa}(b|{\it g});
\label{67}
\ee
where {\it g} is an interaction constant extracted from potential
${\bf V}(r)$ for
Hamiltonians (\ref{8}a,b) respectively. It follows directly
from definition (\ref{1}) using easily verified property of radial
Green function (see Appendix):
\[
G^{-\overline{\zeta}}_{-\kappa 0}(b;r,y)\ =\
-\sigma_{1}G^{\overline{\zeta}}_{\kappa 0}(b;r,y)\sigma_{1}.
\]

\section{\rm Nonlocal and singular interactions}

There is a close connection between nonlocal and singular interaction.
Its become apparent, on the one hand, via constraction of the selfadjoint
extension for nonselfadjoint singular Hamiltonian with the help of nonlocal
interaction \cite{FddBr}, on the other hand, via existence of Hamiltonians
admitting dual interpretation, as nonlocal from one point of view, and as
local but singular from the other one.
We start with observation, that for Hamiltonian (\ref{18}) in spite of
its manifest dependence from momentum, pointing its nonlocality,
the expression for each partial wave has local form:
\begin{equation}
<JlM|2m {\bf U}(\vec{\rm x})|JlM>=U_{\kappa_{\xi}}(r)=
\int_{\mu_{0}}^{\infty} d\nu\
\Sigma_{\kappa_{\xi}}(\nu)\ \frac{e^{-\nu r}}{r}-I^{\Sigma}_{0}
\frac{\delta (r)}{2(2m)^{2}r^{2}};
\label{68}
\end{equation}
with singular behavior at $r \rightarrow 0$:
\begin{equation}
U_{\kappa_{\xi}}(r) \rightarrow I^{\Sigma}_{0}
\frac{(1+\xi)}{(2m)^{2}r^{3}}-
I^{\Sigma}_{0}\frac{\delta(r)}{2(2m)^{2}r^{2}}+
\frac{1}{r}
\left[
I^{\Sigma}_{0}-I^{\Sigma}_{2}\frac{(\kappa_{\xi}+1/2)}{(2m)^{2}}
\right],
\label{70}
\end{equation}
where:
\be
I^{\Sigma}_{n}=\int_{\mu_{0}}^{\infty}d\nu\, \Sigma(\nu)\,\nu^{n};
\label{70*}
\ee
and
\begin{equation}
\Sigma_{\kappa_{\xi}}(\nu)=\Sigma (\nu)+(2m)^{-2}
\left[
\frac{\nu^{2}}{2}\Sigma(\nu)+(1+\kappa_{\xi})\nu
\int_{\mu_{0}}^{\nu} d\gamma \, \Sigma(\gamma)
\right].
\label{69}
\end{equation}
Corresponding Volterrian kernel may be expressed either in the form
like (\ref{50*}) with density (\ref{69}) instead $\Sigma^{(3)}(\nu)$, or in
the form like (\ref{50}) as (clf.(\ref{35*})):
\begin{equation}
{\tt K}_{l_{\xi}J}(u,\varrho)=\tilde{A}^{(1)}(iu,i \varrho)\ K_{l_{\xi}}(u,\varrho)+
\tilde{A}^{(2)}(iu,i \varrho)\ K_{l_{-\xi}}(u,\varrho).
\label{71}
\end{equation}
In the case (\ref{19}) we have for kernel ${\cal K}_{l}(u,\varrho)$
the form (\ref{50*}) with density (\ref{35**}).
It is not difficult to see, that although for considering interactions the
integrals defining auxiliary kernel (\ref{48}),(\ref{49}) do not exist, their
difference has a definite value: $(l=l_{\xi}),$
\begin{equation}
{\tt H}_{lJ}(\varrho,s)-{\tt H}_{lJ}(\varrho,k)=
(k^{2}-s^{2})
\int_{\varrho+\mu_{0}}^{\infty} d\alpha
\frac{{\tt K}_{lJ}(\alpha,\varrho)}{(\alpha^{2}+s^{2})(\alpha^{2}+k^{2})}
\left(
\frac{\varrho}{\alpha}
\right)^{l}.
\label{72}
\end{equation}
That prompt to turn for this case to ansatz (\ref{42}) subtracted in the point
$\varrho=\Lambda$, where $\Lambda \rightarrow \infty $. Repeating for it all
transformations of the previous section with use of subtracted functions
(written for simplicity for case (\ref{19})):
\be
{\cal H}^{\Lambda}_{l}(\varrho,k)=\left(\frac{\Lambda^{2}+k^{2}}
{\Lambda^{2}-\varrho^{2}}\right)
\left[
{\tt H}_{l}(\varrho,k)-{\tt H}_{l}(\Lambda,k)
\right],
\label{73}
\ee
\begin{eqnarray}
{\cal M}^{\Lambda}_{l}(\varrho^{2},k^{2}) & = &
\left[
F_{l}(\Lambda,-ik)-Z_{l}(\Lambda^{2},k^{2})+Z_{l}(-k^{2},k^{2})
\right]
(\Lambda^{2}-\varrho^{2})^{-1} h(\Lambda^{2}),
\nonumber \\
{\cal F}^{\Lambda}_{l}(\varrho,-ik) & = &
\left[
F_{l}(\varrho,-ik)-Z_{l}(\varrho^{2},k^{2})+Z_{l}(-k^{2},k^{2})
\right]
(\Lambda^{2}-\varrho^{2})^{-1} h(\Lambda^{2}),
\nonumber
\end{eqnarray}
where $h(\Lambda^{2})$ is some appropriate function choosing bellow,
we get for limiting value
\be
{\cal F}_{l}(\varrho,-ik)=\lim_{\Lambda \rightarrow \infty}
{\cal F}^{\Lambda}_{l}(\varrho,-ik)
\label{73*}
\ee
Volterra-type equation similar to Schrodinger version \cite{Ktmf} of 
eq.(\ref{53}):  
\begin{equation} 
{\cal F}_{l}(\varrho,-ik)= 
\lim_{\Lambda \rightarrow \infty} \left[ 
{\cal M}^{\Lambda}_{l}(\varrho^{2},k^{2})+ 
\int_{\varrho+\nu_{0}}^{\infty} du 
\frac{ {\cal K}_{l}(u,\varrho) } { (u^{2}+k^{2}) } 
\left( 
\frac{\varrho}{u}
\right)^{l}\
{\cal F}_{l}(u,-ik)
\right].
\label{74}
\end{equation}
For any finite $\Lambda $ its iteration serie convergent under the conditions:
\begin{equation}
I^{\Sigma_{2}}_{0}=0;\qquad |I^{\Sigma_{2}}_{1}|<\infty,
\label{75}
\end{equation}
and lead to solution in familiar form:
\be
{\cal F}_{l}(\varrho,-ik)=\lim_{\Lambda \rightarrow \infty}
\left[
{\cal M}^{\Lambda}_{l}(\varrho^{2},k^{2})
+\int_{\varrho+\nu_{0}}^{\infty} du
\frac{
{\tt a}_{l}(u,\varrho;-k^{2})
      }
{
(u^{2}+k^{2})
}
\left(
\frac{\varrho}{u}
\right)^{l}
{\cal M}^{\Lambda}_{l}(u^{2},k^{2});
\right].
\label{76}
\ee
Here resolvent ${\tt a}_{l}(u,\varrho;-k^{2})$ satisfy to independent from
$\Lambda$ Schrodinger version of equations (\ref{54}) \cite{Ktmf}
with corresponding kernel ${\tt K}_{lJ}(u,\varrho)$ or ${\cal K}_{l}(u,\varrho)$
(clf. substitutions (\ref{35})). Moreover, it admit also
exact factorization of $l,J$ dependence in terms of T-matrix
momentum transfer spectral density, repeating the form of (\ref{56})
for the case (\ref{18}), and for the case (\ref{19}) repeating a simple
form \cite{Ktmf}. It is well known \cite{BrFLS} that for nonlocal
interaction the right hand side of simple normalization condition
(\ref{44}) is replaced to the determinant ${\cal M}_{l}(k^{2})$ which
has the same form (\ref{1}) with Volterrian Green function
${\bf B}_{l0}(k^{2})$ (clf.Appendix)
instead ${\bf G}_{l0}(W)$, and play a role of the measure of nonlocality
\footnote{So, for Schrodinger Hamiltonian with local nonsingular
potential (\ref{9}) we have instead (\ref{52}) the natural choice
$
Z_{l}(\varrho^{2},k^{2})={\cal M}^{\Lambda}_{l}(\varrho^{2},k^{2})\equiv 1.
$
\cite{Ktmf}}.
However such uncertainty of normalization did not affect on scattering
phase.  If one can choose the function $h(\Lambda^{2})$ so, that the limit
(\ref{73*}),(\ref{76}) exist, then the function ${\cal F}_{l}(\varrho,-ik)$
is renormalazed OSJF for nonlocal "potentials" (\ref{18}),(\ref{19})
and the respective limit of
${\cal M}^{\Lambda}_{l}(\varrho^{2},k^{2})$ is renormalazed determinant
${\cal M}_{l}(k^{2})$ (clf.(\ref{73})).

For Schrodinger operator
\footnote{Without loss of generality here we can restrict ourself by the
case $N=3$ \cite{Ktmf}.}
\begin{equation}
\left({\cal L}^{l}_{r}(b)-V(r)\right)=
\partial^{2}_{r}-l(l+1) r^{-2}-b^{2}-V(r)
\label{77*}
\end{equation}
with singular local potential the definition
(\ref{1}) make no sense, and we left only with possibility to define Jost
function as Wronskian \cite{FLS-Ph}:
\begin{equation}
f_{l}(\mp ik)=
\left(
f_{l}(\mp ik,r)\stackrel{\leftrightarrow}
{\partial_{r}} \varphi_{l}(k^{2},r)
\right).
\label{77}
\end{equation}
Here $f_{l}(\mp ik,r), \varphi_{l}(k^{2},r), I_{l}(k^{2},r)$ are
respectively the Jost, regular, and irregular solutions of Schrodinger
equation:
\begin{equation}
\left(
f_{l}(+ik,r)\stackrel{\leftrightarrow}
{\partial_{r}} f_{l}(-ik,r)
\right)=2ik,\quad
\left(
I_{l}(k^{2},r)\stackrel{\leftrightarrow}
{\partial_{r}} \varphi_{l}(k^{2},r)
\right)=1.
\label{78}
\end{equation}
It is not difficult to show \cite{Kds}, that for $Re\ \varrho >|Im\ k|$
independing from type of potential's singularity, this definition of Jost
function is equivalent to representation:
\begin{equation}
f_{l}(\varrho)=(\varrho^{2}+k^{2}) \int_{0}^{\infty} dr\ \
\varphi_{l}(k^{2},r)\ \ f_{l}(\varrho,r),
\label{79**}
\end{equation}
(where r.h.s. really don't depend from $k^{2}$ in this domain), and as
a consequence, it is related with the following OSJF:
\be
f_{l}(\varrho)=\lim_{k \rightarrow i\varrho}
{\cal T}_{l}(\varrho,-ik),
\label{79*}
\ee
\begin{equation}
{\cal T}_{l}(\varrho,-ik)= (\varrho^{2}+k^{2}) \int_{0}^{\infty} dr\ \
\varphi_{l}(k^{2},r)\ \ \chi_{l}(\varrho r),
\label{79}
\end{equation}
which in its turn is well defined by the natural generalization of
Schrodinger version of (\ref{66}) for singular repulsive potential:
\begin{equation}
{\cal T}_{l}(\varrho,-ik)=\ \lim_{r \rightarrow 0}\
\left[
I_{l}(k^{2},r)
\right]^{-1}
J_{l}(\varrho,-ik;r).
\label{80}
\end{equation}
Therefor from the first line of Schrodinger variant of (\ref{64})
\cite{Ktmf,Krc}, we obtain for it a homogeneous Volterra equation:
\be
{\cal T}_{l}(\varrho,b)= \int_{\varrho+\mu_{0}}^{\infty}du\
\frac{K_{l}(u,\varrho)}{(u^{2}-b^{2})}\
{\cal T}_{l}(u,b).
\label{81}
\ee
It is clear that nontrivial solution of such equation originated from
singular behavior of its kernel $K_{l}(u,\varrho)$ for $u\rightarrow \infty$.
Putting in this limit for resolvent:
\begin{equation}
{\tt a}_{l}(u,\varrho;b^{2}) \Rightarrow A_{l}(u,b^{2})\ C_{l}(\varrho,b^{2});
\label{82}
\end{equation}
one can see, that if:
\be
K_{l}(u,\varrho)\,/\,A_{l}(u,b^{2}) \Rightarrow 0;
\label{83}
\ee
then from the second line of the Schrodinger version of resolvent's
equation (\ref{54})
\cite{Ktmf} the equation for $C_{l}(\varrho,b^{2}) $ follows which is
identical to the (\ref{81}). Thus we can identify supposing for a moment
the continuity on $u$:
\begin{equation}
{\cal T}_{l}(\varrho,b)=C_{l}(\varrho,b^{2})=\lim_{u \rightarrow \infty}
{\tt a}_{l}(u,\varrho;b^{2})\ /\ A_{l}(u,b^{2}).
\label{84}
\end{equation}
The function $A_{l}(u,b^{2})$ may be found independently from the first line
of resolvent equation in the limit $u\rightarrow \infty $, where the kernel
(\ref{50*}) $(N=3)$ is changed by its asymptotic degenerate form:
$K_{l}(u,\varrho)\Rightarrow {\cal U}_{l}(u){\cal R}_{l}(\varrho)\equiv
K^{\infty}_{l}(u,\varrho).
$
The asymptotical solution reads:
\[
A_{l}(u,b^{2})={\cal U}_{l}(u)\ exp[O_{l}(u,b^{2})].
\]
\begin{equation}
O_{l}(u,b^{2})=\int^{u}d\alpha
\frac{
K^{\infty}_{l}(\alpha,\alpha)
      }
{(\alpha^{2}-b^{2})}.
\label{85}
\end{equation}
Because the resolvent ${\tt a}_{l}(u,\varrho;b^{2})$ is in generally a
distribution over $u-\varrho$, we need in corresponding integral form
of relation (\ref{84}). It is easy to verify that such form is nothing
but eq.(\ref{76}), multiplied by $\varrho^{-l}$, with the following
choice of regulator function:
\begin{equation}
{\cal M}^{\Lambda}_{l}(\varrho^{2},k^{2})
\Rightarrow
\frac{1}{\Lambda} exp
\left\{
-\frac{1}{\Lambda} exp\left[O_{l}(\varrho,-k^{2})\right]
\right\}
\label{86}
\end{equation}
The own limit of regulator function in that case is zero in accordance with
homogeneous character of eq.(\ref{81}). One can notice, that for
interaction (\ref{18}) the convergence conditions (\ref{75}) lead to
nonsingular behavior at the origin. We want to emphasize, that iteration
series for spectral density (\ref{58}), resolvent (\ref{54}) and OSJS
(\ref{64})
convergent under essentially more weak restrictions on the potential,
than the serie for OSJF. The corresponding estimations \cite{Kds} show,
that the first are entire functions of coupling constant {\it g}
(and angular momentum $l$ or $J_{\rm N}$) for arbitrary potential
considering here, but the second has well-known essential singularity at
${\it g}=0$ for singular potential \cite{FLS-Ph}, and for regular one
it is regular only at $Re\ J_{\rm N}\geq \lambda_{\rm N}$.
So, for singular or nonlocal interactions we may considering eq.(\ref{76}),
(\ref{79*}) as {\bf definition} of OSJF and Jost function respectively.
The question now is only to choose the regulator function
${\cal M}^{\Lambda}_{l}(\varrho^2,k^{2})$ so that corresponding limit should
exist. For singular repulsive potential the answer is given by the
eq.(\ref{86}). For nonlocal interaction (\ref{19}) one can make it choice
only if the manifest form of resolvent ${\tt a}_{l}(u,\varrho;-k^{2})$ is known.

\section{\rm Conclusions}

In this work the generalization of the OSJF method for Dirac Hamiltonian
in arbitrary dimension space is given. New integral representations for
OSJF, OSJS and Jost determinant via T-matrix momentum transfer
spectral density are found, which together with the linear
Volterrian integral equation for it constitute a closed dynamic system,
allowing to calculate all observable quantities
without dealing with eigenvalue problem.

The OSJF for singular and nonlocal potential
are constructed through this spectral density with the help of common
renormalization procedure, which naturally generalizes such
representation for local nonsingular potential.

Author thanks V.M.Leviant and V.S.Otchik for helpful discussions, and
Yu.V. \\Parfenov for his interest in this work.

\section{\rm Appendix}

We used the following manifest form of partial Green functions:
\begin{equation}
G^{\zeta}_{\kappa_{\xi}0}(-ik;r,y)=
\theta(y-r)\ B^{\zeta}_{\kappa_{\xi}0}(-ik;r,y)-
\frac{(-i)^{L_{\xi}}}{\eta^{\zeta}(k)}\,
X^{\zeta}_{\kappa_{\xi}}(-ik,r)\,
\stackrel{\rm (T)}{\phi}\,^{\zeta}_{\kappa_{\xi}0}(k,y);
\label{A1}
\end{equation}
\be
B^{\zeta}_{\kappa 0}(\pm ik;r,y)=
-\stackrel{\rm (T)}{B}\,^{\zeta}_{\kappa 0}(ik;y,r)
=\frac{1}{2i\eta^{\zeta}(k)}
\left[
X^{\zeta}_{\kappa}(ik,r)\stackrel{\rm (T)}{X}\,^{\zeta}_{\kappa}(-ik,y)
-[k \rightarrow -k]
\right];
\label{A2}
\ee
where
$
\kappa=\kappa_{\xi}, L=L_{\xi}, L-\xi=L_{-\xi},
(2\kappa_{\xi}+1)=\xi (2L_{\xi}+1),
$
and free solutions are:
\be
\psi_{j0}(kr)=\left( \frac{\pi kr}{2} \right)^{1/2}
J_{J+\frac{1}{2}}(kr);
\label{A3}
\ee
\[
\phi^{\zeta}_{\kappa_{\xi}0}(k,r)=
\left[
\begin{array}{c}
\psi_{L_{\xi}0}(kr)
\\
\xi \quad \eta^{\zeta}(k) \quad \psi_{L_{-\xi}0}(kr)
\end{array}
\right]=
\frac{1}{2i}\left[(-i)^{L_{\xi}}\,X^{\zeta}_{\kappa_{\xi}}(-ik,r)
-i^{L_{\xi}}\,X^{\zeta}_{\kappa_{\xi}}(ik,r)\right];
\]
\[
X^{\zeta}_{\kappa}(\varrho,r)=
\left[
\begin{array}{c}
\chi_{\kappa}(\varrho r)
\\
i\eta^{\zeta}(i\varrho) \quad \chi_{-\kappa}(\varrho r)
\end{array}
\right]
\equiv
\left[
\begin{array}{c}
\chi_{L}(\varrho r)
\\
i\eta^{\zeta}(i\varrho) \quad \chi_{L-\xi}(\varrho r)
\end{array}
\right];
\]

Relation (\ref{28}) may be obtained \cite{Kds} analogously with it
particular case $l=0$, in \cite{Ktmf} using the relation \cite{Bt-Er}
for the Legendre function of second kind, for $j=l-a_{N}$:
\begin{equation}
S^{a}_{j}(Z)
\equiv
\frac{
e^{-i\pi a}\ Q^{a}_{j}(Z)
      }
{
\left(
Z^{2}-1
\right)^{a/2}
}=\frac{(-1)^{l}}{2^{j+1}\Gamma(j+1)}\left(\frac{d}{dZ}\right)^{l}
\int_{-1}^{1}dt\ \frac{(1-t^{2})^{j}}{Z-t}.
\label{A4}
\end{equation}
The projector onto the states with given $l^{(N)}_{\xi},J_{\rm N}$ reads:
\begin{eqnarray}
&&
\Pi_{\kappa_{\xi}}(\vec{\tau},\vec{\rm v})=
\frac{\xi}{\Omega_{N}}
\left[
C^{\lambda+1}_{l_{\xi}-1}
(\vec{\tau} \cdot \vec{\rm v})\
(\vec{\sigma} \cdot \vec{\tau})
(\vec{\sigma} \cdot \vec{\rm v})^{\dagger}-
C^{\lambda+1}_{l_{-\xi}-1}
(\vec{\tau} \cdot \vec{\rm v})
\right]=
\nonumber \\
&&
=\frac{\xi}{\Omega_{N}}
\left[
\left(
l_{\xi}+\lambda_{N}(1-\xi)
\right)\
C^{\lambda}_{l_{\xi}}
(\vec{\tau} \cdot \vec{\rm v})+
2i\lambda_{N}\xi
(\vec{\tau} \cdot {\Omega} \cdot \vec{\rm v})\
C^{\lambda+1}_{l_{\xi}-1}
(\vec{\tau} \cdot \vec{\rm v})
\right];
\nonumber \\
\label{A5}
\end{eqnarray}
where:
\begin{equation}
\Omega_{jk}=\frac{1}{2i}
\left(
\sigma_{j}\sigma^{\dagger}_{k}-\sigma_{k}\sigma^{\dagger}_{j}
\right);\quad
{\tilde{\Omega}}_{jk}=\frac{1}{2i}
\left(
\sigma^{\dagger}_{j}\sigma_{k}-\sigma^{\dagger}_{k}\sigma_{j}
\right).
\label{A7}
\end{equation}
It is convenient for separation of variables to use
total antisymmetry of tensor $E_{jknl}$ and relation with it:
\[
E_{jknl}=\frac{i}{2}
\left[
\Gamma_{n}
\left\{
\Gamma_{l},\Sigma_{jk}
\right\}
\right];
\]
\begin{equation}
\left\{
\Sigma_{jk},\Sigma_{nl}
\right\}
=2(\delta_{nj}\delta_{kl}-\delta_{lj}\delta_{kn})-
E_{jknl};
\label{A8}
\end{equation}
\[
\Sigma_{jk}=\frac{1}{2i}[\Gamma_{j}\ ,\Gamma_{k}]=
\left(
\begin{array}{cc}
\Omega_{jk} & 0\\
0 & {\tilde \Omega}_{jk}
\end{array}
\right).
\]
Operators of orbital and total angular momentum are defined as:
\begin{equation}
{\bf L}_{jk}=x_{j}{\bf P}_{k}-x_{k}{\bf P}_{j};\qquad
{\bf J}_{jk}={\bf L}_{jk}+\frac{1}{2}\Sigma_{jk}.
\label{A9}
\end{equation}
Introducing the notations:
\begin{equation}
n_{k}=cos \vartheta_{k-1}\prod^{N-1}_{j=k}sin \vartheta_{j};\quad
\vec{\nabla}_{N}=\vec{\rm n}\partial_{r}+
\frac{1}{r}\vec{\partial}_{\vec{\rm n}};
\label{A10}
\end{equation}
the following relations may be checked:
\begin{equation}
\frac{1}{2}({\bf L} \cdot\Omega)=
-(\vec{\sigma} \cdot \vec{\rm n})
(\vec{\sigma}^{\dagger} \cdot \vec{\partial}_{\vec{\rm n}});\qquad
(\vec{\rm n} \cdot \vec{\partial}_{\vec{\rm n}})=0;
\label{A11}
\end{equation}
which together with addition theorem and recurrence relations for
Gegenbauer polinomous \cite{Bt-Er}:
\[
\frac{d}{dz}C^{\lambda}_{l}(z)=
2\lambda C^{\lambda+1}_{l-1}(z);
\]
\begin{equation}
zC^{\lambda+1}_{l-1}(z)-C^{\lambda+1}_{l-1-\xi}(z)=
\frac{\xi}{2\lambda}[l+\lambda(1-\xi)]\
C^{\lambda}_{l}(z);\ \xi=\pm 1;
\label{AA}
\end{equation}
help to verify the following useful properties:
\begin{equation}
\int d \Omega_{N}(\vec{\rm n})\ \
\Pi_{\kappa_{\xi}}(\vec{\omega},\vec{\rm n})\ \
\Pi_{\tau_{\varsigma}}(\vec{\rm n},\vec{\rm v})=
\delta_{\xi \varsigma}\,\delta_{|\kappa|,|\tau|}\
\Pi_{\kappa_{\xi}}(\vec{\omega},\vec{\rm v});
\label{A12}
\end{equation}
\begin{eqnarray}
&\ &
\frac{4\pi}{(2\pi)^{\alpha}}
\sum_{J_{\rm N}=\lambda_{N}}^{\infty}
\sum_{\xi=\pm 1}
S^{a}_{L_{\xi}}(Z)
\left\{
\begin{array}{c}
\Pi_{\kappa_{\xi}}(\vec{\tau},\vec{\rm v})
\\
\Pi_{\kappa_{-\xi}}(\vec{\tau},\vec{\rm v})
\end{array}
\right\}=
\nonumber \\
&\ &
=
\left\{
\begin{array}{c}
{\tt I}
\\
(\vec{\sigma} \cdot \vec{\tau})
(\vec{\sigma} \cdot \vec{\rm v})^{\dagger}
\end{array}
\right\}
\frac{1}{\left[Z-(\vec{\tau} \cdot \vec{\rm v})\right]};\nonumber \\
\label{A13}
\end{eqnarray}
\[
\left[
\frac{N-1}{2}+\frac{1}{2}({\bf L} \cdot {\Omega})
\right]
\Pi_{\kappa_{\xi}}(\vec{\rm n},\vec{\omega})=
-\kappa_{\xi}
\Pi_{\kappa_{\xi}}(\vec{\rm n},\vec{\omega});
\]
\begin{equation}
\left[
\frac{N-1}{2}+\frac{1}{2}({\bf L}\cdot \tilde{\Omega})
\right]
(\vec{\sigma} \cdot \vec{\rm n})^{\dagger}
\Pi_{\kappa_{\xi}}
(\vec{\rm n},\vec{\omega})=
\kappa_{\xi}
(\vec{\sigma} \cdot \vec{\rm n})^{\dagger}
\Pi_{\kappa_{\xi}}(\vec{\rm n},\vec{\omega});
\label{A14}
\end{equation}
\be
\frac{1}{2}({\bf L \cdot \bf L})\,
\Pi_{\kappa_{\xi}}(\vec{\rm n},\vec{\omega})=
l_{\xi}^{(N)}\left(l_{\xi}^{(N)}+2\lambda_{N}\right)\,
\Pi_{\kappa_{\xi}}(\vec{\rm n},\vec{\omega});\quad
\label{A15}
\ee

\end{document}